\documentclass[twocolumn]{aastex631}

\usepackage{xspace}

\newcommand{\snia}{SN~Ia\xspace}
\newcommand{\sneia}{SNe~Ia\xspace}
\newcommand{\hst}{\emph{HST}\xspace}
\newcommand{\name}{SN~2011fe\xspace}
\newcommand{\filter}{\hbox{\ensuremath{F555W}}\xspace}
\newcommand{\tmax}{\hbox{\ensuremath{t_{\rm{max}}}}\xspace}

\newcommand{\Teff}{\hbox{\ensuremath{T_{\rm eff}}}\xspace}
\newcommand{\Lstar}{\hbox{\ensuremath{\rm L_\star}}\xspace}
\newcommand{\Msun}{\hbox{\ensuremath{\rm M_\odot}}\xspace}
\newcommand{\Lsun}{\hbox{\ensuremath{\rm L_\odot}}\xspace}

\begin{document}

\title{The HST Non-Detection of SN~Ia~2011fe 11.5~yr After Explosion\\Further Restricts Single-Degenerate Progenitor Systems}

\author{M. A. Tucker}
\altaffiliation{CCAPP Fellow}
\affiliation{Center for Cosmology and Astroparticle Physics, The Ohio State University, 191 West Woodruff Ave, Columbus, OH, USA}
\affiliation{Department of Astronomy, The Ohio State University, 140 West 18th Avenue, Columbus, OH, USA}
\affiliation{Department of Physics, The Ohio State University, 191 West Woodruff Ave, Columbus, OH, USA}
\email{tucker.957@osu.edu}

\author{B. J. Shappee}
\affiliation{Institute for Astronomy, University of Hawaii, 2680 Woodlawn Drive, Honolulu HI 96822, USA}


\begin{abstract}

We present deep \textit{Hubble Space Telescope} imaging of the nearby Type Ia supernova (SN Ia) 2011fe obtained 11.5~yr after explosion. No emission is detected at the SN location to a $1\sigma$ ($3\sigma$) limit of ${F555W > 30.2 \;(29.0)}$~mag, or equivalently $M_V > 1.2 \;(-0.1)$~mag, neglecting the distance uncertainty to M101. We constrain the presence of donor stars impacted by the SN ejecta with the strictest limits thus far on compact (i.e., $\log \,g \gtrsim 4$) companions. H-rich zero-age main-sequence companions with masses $\geq 2~\Msun$ are excluded, a significant improvement upon the pre-explosion imaging limit of $\approx 5~\Msun$. Main-sequence He stars with masses $\geq 1.0~\Msun$ and subgiant He stars with masses $\leq 0.8~\Msun$ are also disfavored by our late-time imaging. Synthesizing our limits on post-impact donors with previous constraints from pre-explosion imaging, early-time radio and X-ray observations, and nebular-phase spectroscopy, essentially all formation channels for \name invoking a non-degenerate donor star at the time of explosion are unlikely. 

\end{abstract}

\keywords{Type Ia supernovae (1728), Stellar mass loss (1613), Helium-rich stars (715), Interacting binary stars (801)}


\section{Introduction} \label{sec:intro}

The single-degenerate scenario for producing Type Ia supernovae (\sneia) invokes a non-degenerate donor star undergoing Roche Lobe overflow (RLOF) to transfer mass onto the white dwarf (WD) until reaching the necessary central densities for carbon ignition \citep{whelan1973, nomoto1982}. Mass transfer via RLOF restricts the distance between the WD and the donor to $a\lesssim 3R_\star$ for semi-major axis $a$ and companion radius $R_\star$ \citep{eggleton1983} resulting in the \snia ejecta impacting the companion within moments of the explosion (e.g., \citealp{wheeler1975}; see \citealp{liu2023} for a recent review of observables). The impact deposits energy into the envelope of the companion \citep[e.g., ][]{marietta2000, boehner2017}, heating and expanding the outer layers and becoming overluminous for $\sim 10^3$~yr \citep[e.g., ][]{pod2003, shappee2013}.

Searches for these post-impact donors have mostly been confined to nearby \snia remnants that are $\lesssim 1000$~yr old when the donor is still expected to be significantly overluminous. Some tentative candidates have been reported for Galactic and Magellanic Clouds \snia remnants \citep[e.g., ][]{ruizlapuente2004, ihara2007, li2019} but no unambiguous surviving donor stars have been identified thus far (e.g., \citealp{schaefer2012, kerzendorf2013, pagnotta2015, kerzendorf2018, shields2023}; see \citealp{ruizlapuente2019} for a recent review). However, this experiment is observationally difficult due to the cost of obtaining deep spectroscopic follow-up for many targets and a limited number of nearby and young \snia remnants. 

An alternative method for identifying post-impact companion stars is to obtain deep imaging for nearby \sneia many years after explosion once the SN has sufficiently faded. However, this is only possible for the most nearby ($\lesssim 10$~Mpc) \sneia due to crowding and faintness constraints. \citet{do2021} recently searched for a post-impact donor star for SN~1972E but did not find any candidates in \hst imaging $\approx 33$~yr after explosion. 

In this Letter, we present deep \emph{Hubble Space Telescope} (\hst) imaging of the nearby and well-studied \snia 2011fe \citep{nugent2011} $\approx 11.5$~yr after explosion to constrain the presence of post-impact companions. We adopt a distance to M101 of $6.4\pm0.4$~Mpc \citep[$\mu = 29.03\pm0.14$~mag; ][]{shappee2011} to ease comparisons with previous studies of \name. We correct for the small amount of Milky Way reddening ($E(B-V)_{\rm MW} = 0.008$~mag; \citealp{schlafly2011}) but do not correct for any host-galaxy reddening toward \name as it is negligible \citep{patat2013}. The date of explosion is MJD~55797 \citep{pereira2013}.

\section{HST observations}

\begin{table}
    \centering
    \begin{tabular}{cccc}
    \hline\hline
    Phase [d]$^a$ & $10^{-2}$ counts/s & Vega mag. & $M_V$~[mag] \\\hline
    1125 & $307\pm7$ & $24.47\pm0.03$ & $-4.56$\\
    1304 & $127\pm4$ & $25.45\pm0.04$ & $-3.58$\\
    1403 & $102\pm2$ & $25.71\pm0.03$ & $-3.32$\\
    1623 & $43.3\pm 5.3$ & $26.63\pm0.13$ & $-2.40$ \\
    1840 & $21.1\pm 3.5$ & $27.43\pm0.18$ & $-1.60$ \\
    2104 & $18.0\pm 3.1$ & $27.57\pm0.19$ & $-1.46$\\
    2389 & $9.08\pm 2.93$ & $28.34\pm0.35$ & $-0.69$ \\
    4190 & $-0.29\pm 1.81$ & $>28.96^b$ & $> -0.07^b$\\
    \hline\hline
    \end{tabular}
    \caption{\hst/WFC3 $F555W$ photometry of \name.  Quoted uncertainties are $1\sigma$. \\
    $^a$Relative to \tmax ($\approx t_{\rm exp} + 18$). \\$^b3\sigma$ limit.}
    \label{tab:obs}
\end{table}

We obtain deep imaging of \name using the \hst Wide Field Camera 3 (WFC3) UVIS module. Imaging was only conducted in the \filter filter due to the expected faintness of \name at these epochs. Six images were obtained on UT 2023-03-04 (MJD 60007.55\footnote{Weighted by the exposure time for each observation.}) corresponding to 4208~d (11.5~yr) after explosion with a total exposure time of $8400$~s. To characterize the long-term evolution of \name we include previous \hst observations published by \citet{shappee2017} and \citet{tucker2022b} spanning $\approx 1100-2400$~d after explosion. 

\begin{figure*}
    \centering
    \includegraphics[width=\linewidth]{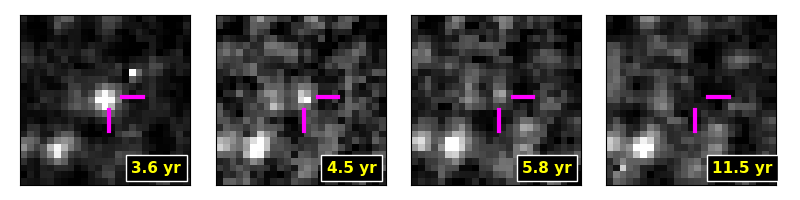}
    \caption{Image cutouts centered on \name for several \hst epochs. The time between explosion and observation is denoted in the lower-right corner of each panel.}
    \label{fig:cutouts}
\end{figure*}

\begin{figure}
    \centering
    \includegraphics[width=\linewidth]{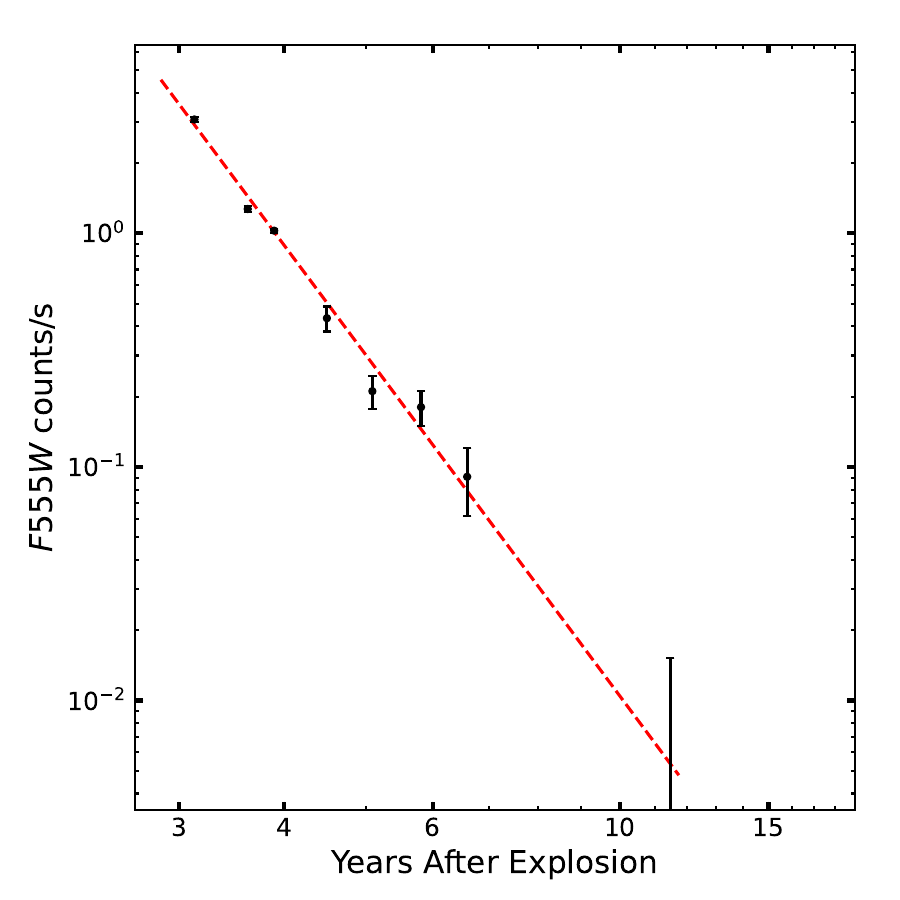}
    \caption{$F555W$ light curve of \name provided in Table~\ref{tab:obs}. The red dashed line shows a simple power-law fit with $f\propto t^{-5}$ highlighting the steadily declining flux. The last observation is not included when fitting the power-law model.}
    \label{fig:f555w}
\end{figure}

Individual images were aligned with \textsc{TweakReg} and combined with \textsc{AstroDrizzle} \citep{astrodrizzle} before performing point spread function (PSF) fitting photometry with \textsc{Dolphot} \citep{dolphin2000, dolphin2016}. Time-dependent filter zeropoints are taken from the WFC3 headers \citep{calamida2023}. All images are analyzed simultaneously to ensure the location of \name remains consistent across images. \name is formally undetected in the last epoch of \hst imaging obtained 11.5~yr after the explosion so we validate the reported uncertainties by checking the photometry of nearby sources. The \filter light curve of \name is provided in Table~\ref{tab:obs}. Non-detection limits are computed via $m>-2.5\log_{10}(3\sigma_f) + z$ for a given flux uncertainty $\sigma_f$ in counts/s and image zeropoint $z$ in magnitudes. Fig.~\ref{fig:cutouts} shows cutouts around \name for several \hst epochs and the long-term \filter evolution is shown in Fig.~\ref{fig:f555w}.

\section{Non-Detection of A Post-impact Companion}\label{sec:impact}

We compare the $F555W$ light curve of \name to models of post-impact He (\S\ref{subsec:impact.He}) and H-rich (\S\ref{subsec:impact.H}) companions. The \textsc{stsynphot} software \citep{stsynphot} is used to impute synthetic \hst $F555W$ magnitudes (in the Vega system) using the \Teff and \Lstar values taken from the models. When referring to the stellar properties of different models throughout the following sections, we always report the values at the moment of SN explosion (i.e., after mass transfer but before the ejecta impact) to avoid confusion.

\subsection{He-star Donors}\label{subsec:impact.He}

Models of the post-impact evolution for He-burning donors are taken from \citeauthor{pan2013} (\citeyear{pan2013}, models HeWDa--d; see also \citealp{pan2010, pan2012b}) and \citeauthor{liu2022} (\citeyear{liu2022}, models He01r and He02r; see also \citealp{liu2013}) which use the binary evolution results of \citet{wang2009} to construct their initial binary systems. We also include the simulations of \citet{liu2021} for a double-detonation explosion triggered by accretion from a low-mass MS He-star (models A, B, and C). 

\begin{figure*}
    \centering
    \includegraphics[width=\linewidth]{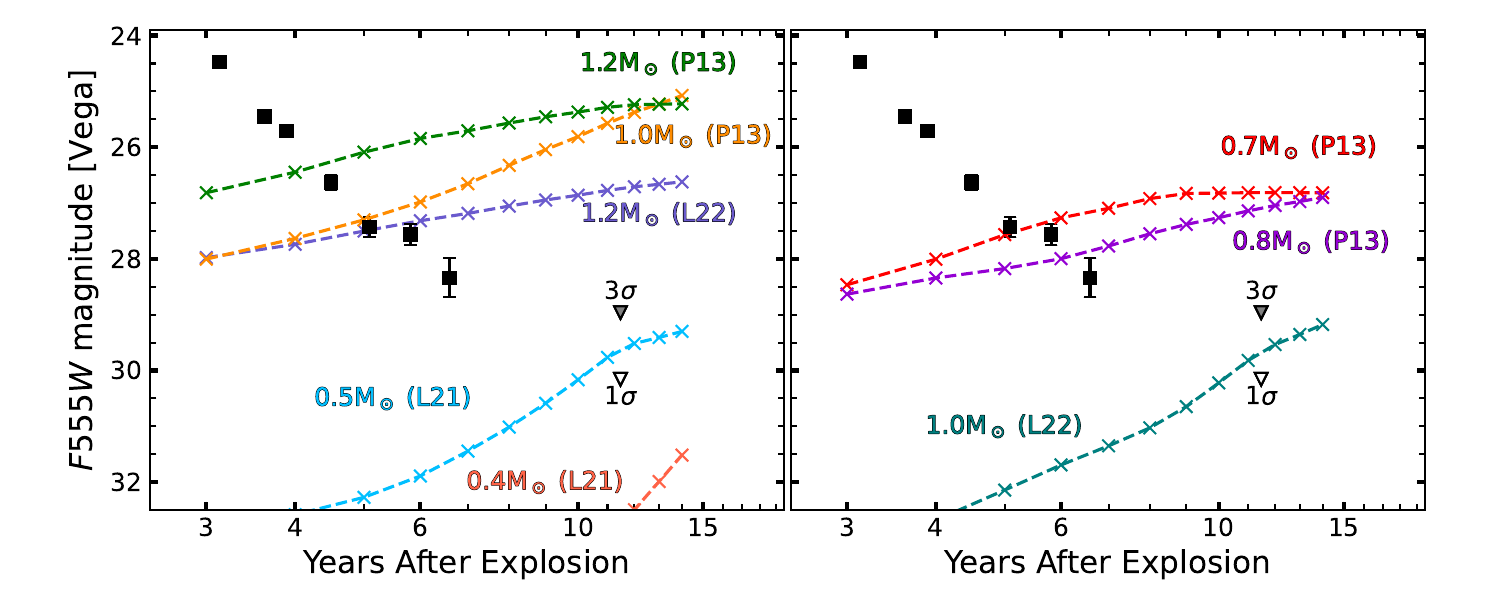}
    \caption{Comparison between the light curve of \name (black squares) with models of post-impact He-star companions (colored lines). The inverted triangles show the $1\sigma$ (open triangle) and $3\sigma$ (filled triangle) non-detection limits. The uncertainty in the distance modulus is $\approx 0.14$~mag. \textit{Left}: Main-sequence He-star companions from \citeauthor{pan2013} (\citeyear{pan2013}, P13), \citeauthor{liu2021} (\citeyear{liu2021}, L21), and \citeauthor{liu2022} (\citeyear{liu2022}, L22). \textit{Right}: Similar to the left panel except for SG He-star companions.}
    \label{fig:He-models}
\end{figure*}

Fig.~\ref{fig:He-models} shows that most He-star models are incompatible with the observed $F555W$ light curve of \name. The three lowest-mass He donors ($0.30~\rm M_\odot$, $0.40~\rm M_\odot$, $0.50~\rm M_\odot$) from \citet{liu2021} would not be detected. Thus, we disfavor MS He-burning donors with masses $\gtrsim 1.0~\rm M_\odot$ because the thermal timescale decreases with increasing mass. The inverse is true for the subgiant (SG)  models, where more massive donors produce more extended envelopes and thus increase the thermalization timescale. The $1.0~\Msun$ model from \citet{liu2022} would not be detected in our \hst observations so we disfavor $\leq 0.8~\Msun$ SG He-stars.

\subsection{H-rich Donors}\label{subsec:impact.H}

\begin{figure*}
    \centering
    \includegraphics[width=\linewidth]{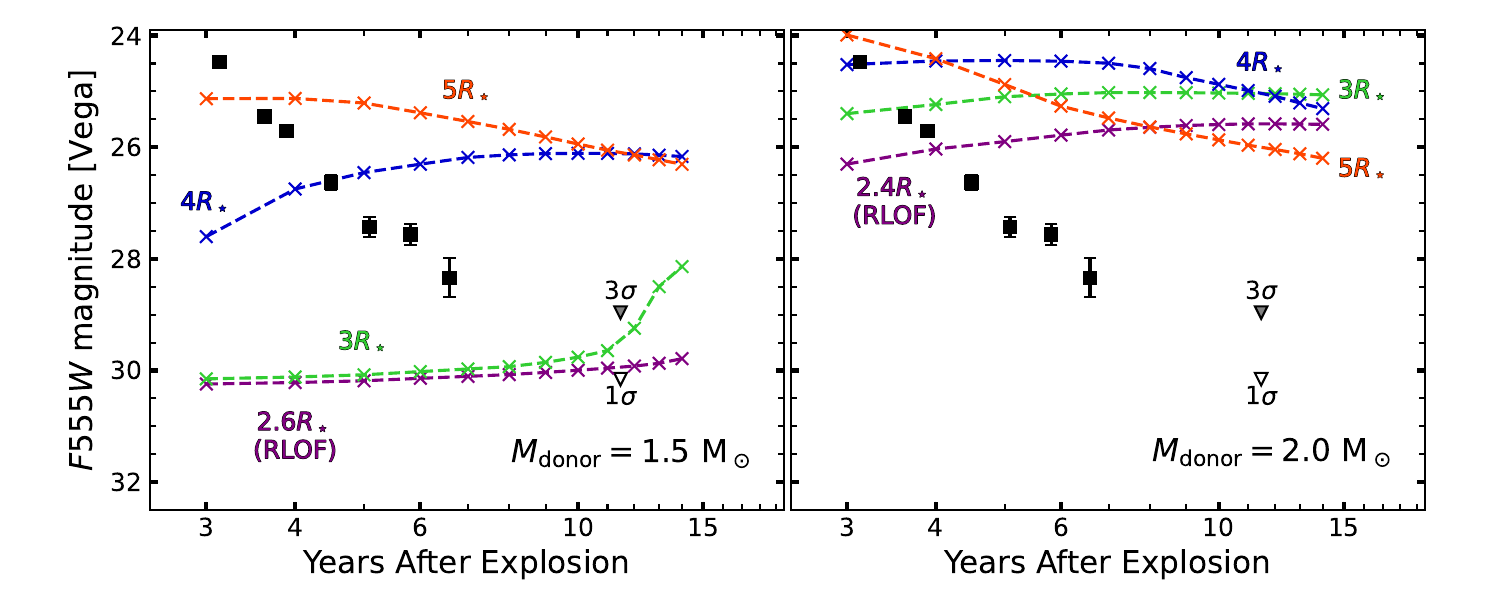}
    \caption{Comaprison between the light curve of \name (black squares) with the post-impact H-rich MS companions of \citet{rau2022} with masses of $M_{\rm donor} = 1.5~\Msun$ (left) and $M_{\rm donor} = 2~\Msun$ (right) for different binary separations. The uncertainty in the distance is $\approx 0.14$~mag.}
    \label{fig:H-models}
\end{figure*}

Unlike He-stars, most H-burning MS donors remain undetectable ($F555W \gtrsim 32$~mag) over the observational baseline \citep{pan2012b, rau2022}. This is attributed to H-rich stars having larger radii at a given mass compared to He stars and thus respond slower and reach lower maximum luminosities \citep[e.g., ][]{pan2012b, pan2013}. However, there are a few cases where H-rich donors can be assessed with the \hst non-detection. 

\citet{rau2022} compute models for zero-age main-sequence (ZAMS) companions assuming RLOF and extend the simulations to binary systems with companions beyond the canonical RLOF separation. While such systems are somewhat contrived due to fine-tuning the mass transfer and time of ignition for the C/O WD, the larger separations produce shallower energy deposition which, in turn, corresponds to increased post-impact luminosities and shorter thermalization timescales (e.g., Fig.~8 in \citealp{rau2022}). 

Fig.~\ref{fig:H-models} shows that the $2~\Msun$ donor from \citet{rau2022} is disfavored for all binary separations. The 1.5~\Msun model would have been marginally detected ($\approx 2\sigma$) at $a\gtrsim 3 R_\star$ and undetected if the system was in RLOF. The two lower-mass models from \citet{rau2022}, $0.8~\Msun$ and $1~\Msun$, are not constrained by our observations regardless of separation due to their low peak luminosities ($\lesssim 10~\Lsun$). 

The MS models computed by \citet{pan2012b} with masses of $\approx 1.2-1.9~\Msun$ remain undetectable for another century or more. The differences between these models and the similar-mass models computed by \citet{rau2022} are attributed to the density structure of the donor star at the time of impact. \citet{pan2012b} model the full binary evolution, including mass transfer, when constructing their donor stars whereas \citet{rau2022} assume ZAMS stars to facilitate parameter exploration and dependencies on SN properties. These differences will be discussed further below. 

\section{Discussion}\label{sec:discuss}

The compressibility of the stellar envelope determines the thermal response timescale \citep{pan2012b}. Stars with high surface gravities confine the energy deposition to the outermost layers whereas the energy is deposited deeper into the stellar interior for low $\log g$ donors (cf. Fig.~5 from \citealp{rau2022}). This work constrains a complimentary parameter space to other searches for non-degenerate donor stars in \name as we are most sensitive to high $\log g$ companions. 

All but the lowest-mass MS He donors are excluded by our late-time imaging of \name. The $0.5~\Msun$ model from \citet{liu2021} would not be detected so $\geq 1.0~\Msun$ MS He donors are disfavored. For an assumed accreted He mass of $\approx 0.05~\Msun$ and an accretion efficiency of 50\%\footnote{The adopted 50\% accretion efficiency is purely instructive as the true accretion efficiency depends sensitively on the mass transfer rate \citep[e.g., ][]{piersanti2014, wu2017}. He shells with masses $\gtrsim 0.05~\Msun$ produce distinct early-time signatures \citep[e.g., ][]{polin2019, collins2022} that are not observed in \name.}, our results restrict MS He donors to $\lesssim 1.1~\Msun$ at the onset of RLOF. While such systems are observed in the Milky Way (i.e., AM CVn binaries) they are likely the progenitors of faint and spectroscopically peculiar \sneia \citep[e.g., ][]{bildsten2007, neunteufel2019} instead of normal \sneia such as \name.

The SG He donors follow an opposing trend, with more massive companions being harder to detect due to the increasing mass of the envelope which increases the heating depth and thermalization timescale. SG He donors with masses $\gtrsim 0.9~\Msun$ (again assuming the mass of the He shell is $\lesssim 0.05~\Msun$ and an accretion efficiency of 50\%) are disfavored by our late-time imaging. Higher-mass SG He donors are inconsistent with pre-explosion imaging \citep{li2011, graur2014} so binaries with SG He stars are disfavored for \name. 

H-rich donors are less constrained by our latest epoch of \hst imaging but complement existing non-detections of H-rich companions. All ZAMS H-rich donors with masses $\gtrsim 2~\Msun$ are disfavored, representing a distinct improvement on the pre-explosion limit of $\approx 5~\Msun$ \citep{li2011}. $1.5~\Msun$ donors beyond $4~\rm R_\star$ are also disfavored but smaller separations would not be detected, and $\leq 1~\Msun$ donors at $< 5~R_\star$ are unconstrained by the late-time \hst imaging due to their low post-impact luminosity ($\lesssim 10~\Lsun$). 

However, these same H-rich ZAMS donors are already disfavored along separate lines of evidence. The impacting ejecta will strip or ablate material off the surface of the donor and heating from the radioactively-decaying ejecta will produce strong H emission lines in the nebular phase \citep{mattila2005, botyanszki2018, dessart2020}. H emission is not seen in the spectra of \name \citep{shappee2013a} out to 1000~d after explosion \citep{graham2015, taubenberger2015} and the formal limits on unbound donor material are $\lesssim 10^{-3}~\Msun$ (e.g., Fig.~6 in \citealp{tucker2022a}). All post-impact ZAMS donors not excluded by our \hst observations produce $\gtrsim 0.05~\Msun$ of unbound material, inconsistent with nebular-phase observations \citep{shappee2013a, lundqvist2015}. Our results also restrict `spin-up/spin-down' scenarios \citep[e.g., ][]{justham2011, distefano2011} due to smaller companion radii increasing the surface gravity, decreasing the heating depth, and producing brighter companions after impact (see Fig.~\ref{fig:H-models} and \citealp{rau2022}). Thus, H-rich donors are also disfavored for \name. 

One potential caveat for future observational and theoretical work on post-impact donor stars is the effect of the donor star structure. We qualitatively compare the H-rich $2~\Msun$ ZAMS model at RLOF from \citet{rau2022} to Model B from \citet{pan2012b} with a mass of $1.92~\Msun$ at the time of the explosion. While the right panel of Fig.~\ref{fig:H-models} shows the former is disfavored by our observations, the latter is unconstrained by our observations ($F555W > 32$~mag). The difference between these models is the density profile of the companion at the moment of impact, as the models of \citet{pan2012b} include mass transfer prior to explosion instead of adopting a ZAMS density profile. The mass transfer reduces the envelope density compared to a ZAMS star with identical mass. This is supported by the higher amount of unbound mass in the \citet{pan2012b} evolved model ($\approx 15\%$) compared to the \citet{rau2022} ZAMS model ($\approx 10\%$). This qualitative comparison highlights the differences in post-impact evolution with and without including the effects of mass loss on the donor star structure. 

The $\approx 1$~mag difference between the $1.2~\Msun$ MS He-star models of \citet{pan2013} and \citet{liu2022} seen in Fig.~\ref{fig:He-models} further highlights the effect of mass-transfer on the donor structure. Despite the donors having similar mass at the moment the WD explodes, they began with different masses\footnote{The initial He-donor masses for the $\approx 1.2~\Msun$ models of \citet{pan2013} and \citet{liu2022} are $1.8~\Msun$ and $1.55~\Msun$, respectively, at the onset of RLOF. } and experienced different mass-transfer histories. The observed difference in synthetic $F555W$ (and the underlying $L_\star$ estimates) are driven by inherent differences in the donor's internal density profile. Thus, all constraints on post-impact companions depend on the underlying assumptions used to construct the density profile of the companion. We encourage future simulation efforts to explore the dependence on different density profiles produced by realistic mass-transfer histories. 

It is worth noting that some constraints on a double-degenerate system can be derived, assuming a companion WD survives the explosion as in the `D6' scenario \citep{shen2018}. \citet{shen2017} show that winds can be driven from the WD surface by pollution from radioactive species in the SN ejecta. However, the primary issue with this constraining these models with the \hst observations of \name is the high temperatures ($T_{\rm eff} > 10^5$~K) shifting the majority of the emission to UV wavelengths. The UV photons likely cannot escape the Fe-rich SN ejecta due to extensive neutral and singly-ionized Fe transitions at these wavelengths \citep[e.g., ][]{iron1, iron2}. This will likely cause the observed radiation to deviate strongly from a blackbody and complicates reliable comparison to observations. Additionally, the excess luminosity from the polluted WD fades on similar timescales as the radioactively-decaying ejecta. Thus, one must simultaneously fit the isotopic ratios produced during the explosion \citep[e.g., ][]{tucker2022b} and the emission contribution from the surviving WD.
This should be possible once radiative-transfer calculations can be incorporated into the \citet{shen2017} models.

\name has been a boon for understanding the complex physics governing \snia explosions. These observations, at 11.5~yr after explosion, provide the strongest limits on He-rich donors in addition to further disfavoring H-rich donors. Assessing our new imaging in conjunction with prior limits on the progenitor system of \name \citep{li2011, nugent2011, bloom2012, margutti2012, chomiuk2012, brown2012}, almost all non-degenerate donor stars are observationally disfavored. The remaining scenarios that cannot be formally excluded, such as very low-mass ($\lesssim 0.6~\Msun$) He donors, are disfavored by rate arguments \citep[e.g., ][]{bildsten2007, neunteufel2019} given that \name is a quintessential example of the \snia population. 

\vspace{0.25in}

\facility{HST (WFC3/UVIS)}

\vspace{0.25in}

\emph{Software}: astropy \citep{astropy}, numpy \citep{numpy}, matplotlib \citep{matplotlib}, pandas \citep{pandas}, scipy \citep{scipy}, lmfit \citep{lmfit}, synphot \citep{synphot}

\section*{Acknowledgements}

We thank Michelle Tucker, Jen van Saders, and Ken Shen for useful discussions.


\bibliography{ref}{}

\begin{thebibliography}{}
\expandafter\ifx\csname natexlab\endcsname\relax\def\natexlab#1{#1}\fi
\providecommand{\url}[1]{\href{#1}{#1}}
\providecommand{\dodoi}[1]{doi:~\href{http://doi.org/#1}{\nolinkurl{#1}}}
\providecommand{\doeprint}[1]{\href{http://ascl.net/#1}{\nolinkurl{http://ascl.net/#1}}}
\providecommand{\doarXiv}[1]{\href{https://arxiv.org/abs/#1}{\nolinkurl{https://arxiv.org/abs/#1}}}

\bibitem[{{Astropy Collaboration} {et~al.}(2022){Astropy Collaboration}, {Price-Whelan}, {Lim}, {Earl}, {Starkman}, {Bradley}, {Shupe}, {Patil}, {Corrales}, {Brasseur}, {N{\"o}the}, {Donath}, {Tollerud}, {Morris}, {Ginsburg}, {Vaher}, {Weaver}, {Tocknell}, {Jamieson}, {van Kerkwijk}, {Robitaille}, {Merry}, {Bachetti}, {G{\"u}nther}, {Aldcroft}, {Alvarado-Montes}, {Archibald}, {B{\'o}di}, {Bapat}, {Barentsen}, {Baz{\'a}n}, {Biswas}, {Boquien}, {Burke}, {Cara}, {Cara}, {Conroy}, {Conseil}, {Craig}, {Cross}, {Cruz}, {D'Eugenio}, {Dencheva}, {Devillepoix}, {Dietrich}, {Eigenbrot}, {Erben}, {Ferreira}, {Foreman-Mackey}, {Fox}, {Freij}, {Garg}, {Geda}, {Glattly}, {Gondhalekar}, {Gordon}, {Grant}, {Greenfield}, {Groener}, {Guest}, {Gurovich}, {Handberg}, {Hart}, {Hatfield-Dodds}, {Homeier}, {Hosseinzadeh}, {Jenness}, {Jones}, {Joseph}, {Kalmbach}, {Karamehmetoglu}, {Ka{\l}uszy{\'n}ski}, {Kelley}, {Kern}, {Kerzendorf}, {Koch}, {Kulumani}, {Lee}, {Ly}, {Ma}, {MacBride}, {Maljaars}, {Muna}, {Murphy}, {Norman},
  {O'Steen}, {Oman}, {Pacifici}, {Pascual}, {Pascual-Granado}, {Patil}, {Perren}, {Pickering}, {Rastogi}, {Roulston}, {Ryan}, {Rykoff}, {Sabater}, {Sakurikar}, {Salgado}, {Sanghi}, {Saunders}, {Savchenko}, {Schwardt}, {Seifert-Eckert}, {Shih}, {Jain}, {Shukla}, {Sick}, {Simpson}, {Singanamalla}, {Singer}, {Singhal}, {Sinha}, {Sip{\H{o}}cz}, {Spitler}, {Stansby}, {Streicher}, {{\v{S}}umak}, {Swinbank}, {Taranu}, {Tewary}, {Tremblay}, {de Val-Borro}, {Van Kooten}, {Vasovi{\'c}}, {Verma}, {de Miranda Cardoso}, {Williams}, {Wilson}, {Winkel}, {Wood-Vasey}, {Xue}, {Yoachim}, {Zhang}, {Zonca}, \& {Astropy Project Contributors}}]{astropy}
{Astropy Collaboration}, {Price-Whelan}, A.~M., {Lim}, P.~L., {et~al.} 2022, \apj, 935, 167, \dodoi{10.3847/1538-4357/ac7c74}

\bibitem[{{Avila} {et~al.}(2015){Avila}, {Hack}, {Cara}, {Borncamp}, {Mack}, {Smith}, \& {Ubeda}}]{astrodrizzle}
{Avila}, R.~J., {Hack}, W., {Cara}, M., {et~al.} 2015, in Astronomical Society of the Pacific Conference Series, Vol. 495, Astronomical Data Analysis Software an Systems XXIV (ADASS XXIV), ed. A.~R. {Taylor} \& E.~{Rosolowsky}, 281, \dodoi{10.48550/arXiv.1411.5605}

\bibitem[{{Bautista}(1997)}]{iron2}
{Bautista}, M.~A. 1997, \aaps, 122, 167, \dodoi{10.1051/aas:1997327}

\bibitem[{{Bildsten} {et~al.}(2007){Bildsten}, {Shen}, {Weinberg}, \& {Nelemans}}]{bildsten2007}
{Bildsten}, L., {Shen}, K.~J., {Weinberg}, N.~N., \& {Nelemans}, G. 2007, \apjl, 662, L95, \dodoi{10.1086/519489}

\bibitem[{{Bloom} {et~al.}(2012){Bloom}, {Kasen}, {Shen}, {Nugent}, {Butler}, {Graham}, {Howell}, {Kolb}, {Holmes}, {Haswell}, {Burwitz}, {Rodriguez}, \& {Sullivan}}]{bloom2012}
{Bloom}, J.~S., {Kasen}, D., {Shen}, K.~J., {et~al.} 2012, \apjl, 744, L17, \dodoi{10.1088/2041-8205/744/2/L17}

\bibitem[{{Boehner} {et~al.}(2017){Boehner}, {Plewa}, \& {Langer}}]{boehner2017}
{Boehner}, P., {Plewa}, T., \& {Langer}, N. 2017, \mnras, 465, 2060, \dodoi{10.1093/mnras/stw2737}

\bibitem[{{Boty{\'a}nszki} {et~al.}(2018){Boty{\'a}nszki}, {Kasen}, \& {Plewa}}]{botyanszki2018}
{Boty{\'a}nszki}, J., {Kasen}, D., \& {Plewa}, T. 2018, \apjl, 852, L6, \dodoi{10.3847/2041-8213/aaa07b}

\bibitem[{{Brown} {et~al.}(2012){Brown}, {Dawson}, {de Pasquale}, {Gronwall}, {Holland}, {Immler}, {Kuin}, {Mazzali}, {Milne}, {Oates}, \& {Siegel}}]{brown2012}
{Brown}, P.~J., {Dawson}, K.~S., {de Pasquale}, M., {et~al.} 2012, \apj, 753, 22, \dodoi{10.1088/0004-637X/753/1/22}

\bibitem[{{Calamida} {et~al.}(2022){Calamida}, {Bajaj}, {Mack}, {Marinelli}, {Medina}, {Pidgeon}, {Kozhurina-Platais}, {Shanahan}, \& {Som}}]{calamida2023}
{Calamida}, A., {Bajaj}, V., {Mack}, J., {et~al.} 2022, \aj, 164, 32, \dodoi{10.3847/1538-3881/ac73f0}

\bibitem[{{Chomiuk} {et~al.}(2012){Chomiuk}, {Soderberg}, {Moe}, {Chevalier}, {Rupen}, {Badenes}, {Margutti}, {Fransson}, {Fong}, \& {Dittmann}}]{chomiuk2012}
{Chomiuk}, L., {Soderberg}, A.~M., {Moe}, M., {et~al.} 2012, \apj, 750, 164, \dodoi{10.1088/0004-637X/750/2/164}

\bibitem[{{Collins} {et~al.}(2022){Collins}, {Gronow}, {Sim}, \& {R{\"o}pke}}]{collins2022}
{Collins}, C.~E., {Gronow}, S., {Sim}, S.~A., \& {R{\"o}pke}, F.~K. 2022, \mnras, 517, 5289, \dodoi{10.1093/mnras/stac2665}

\bibitem[{{Dessart} {et~al.}(2020){Dessart}, {Leonard}, \& {Prieto}}]{dessart2020}
{Dessart}, L., {Leonard}, D.~C., \& {Prieto}, J.~L. 2020, \aap, 638, A80, \dodoi{10.1051/0004-6361/202037854}

\bibitem[{{Di Stefano} {et~al.}(2011){Di Stefano}, {Voss}, \& {Claeys}}]{distefano2011}
{Di Stefano}, R., {Voss}, R., \& {Claeys}, J.~S.~W. 2011, \apjl, 738, L1, \dodoi{10.1088/2041-8205/738/1/L1}

\bibitem[{{Do} {et~al.}(2021){Do}, {Shappee}, {De Cuyper}, {Tonry}, {Hunt}, {Schweizer}, {Phillips}, {Burns}, {Beaton}, \& {Hainaut}}]{do2021}
{Do}, A., {Shappee}, B.~J., {De Cuyper}, J.-P., {et~al.} 2021, \mnras, 508, 3649, \dodoi{10.1093/mnras/stab2660}

\bibitem[{{Dolphin}(2016)}]{dolphin2016}
{Dolphin}, A. 2016, {DOLPHOT: Stellar photometry}, Astrophysics Source Code Library, record ascl:1608.013.
\newblock \doeprint{1608.013}

\bibitem[{{Dolphin}(2000)}]{dolphin2000}
{Dolphin}, A.~E. 2000, \pasp, 112, 1383, \dodoi{10.1086/316630}

\bibitem[{{Eggleton}(1983)}]{eggleton1983}
{Eggleton}, P.~P. 1983, \apj, 268, 368, \dodoi{10.1086/160960}

\bibitem[{{Graham} {et~al.}(2015){Graham}, {Nugent}, {Sullivan}, {Filippenko}, {Cenko}, {Silverman}, {Clubb}, \& {Zheng}}]{graham2015}
{Graham}, M.~L., {Nugent}, P.~E., {Sullivan}, M., {et~al.} 2015, \mnras, 454, 1948, \dodoi{10.1093/mnras/stv1888}

\bibitem[{{Graur} {et~al.}(2014){Graur}, {Maoz}, \& {Shara}}]{graur2014}
{Graur}, O., {Maoz}, D., \& {Shara}, M.~M. 2014, \mnras, 442, L28, \dodoi{10.1093/mnrasl/slu052}

\bibitem[{{Harris} {et~al.}(2020){Harris}, {Millman}, {van der Walt}, {Gommers}, {Virtanen}, {Cournapeau}, {Wieser}, {Taylor}, {Berg}, {Smith}, {Kern}, {Picus}, {Hoyer}, {van Kerkwijk}, {Brett}, {Haldane}, {del R{\'\i}o}, {Wiebe}, {Peterson}, {G{\'e}rard-Marchant}, {Sheppard}, {Reddy}, {Weckesser}, {Abbasi}, {Gohlke}, \& {Oliphant}}]{numpy}
{Harris}, C.~R., {Millman}, K.~J., {van der Walt}, S.~J., {et~al.} 2020, \nat, 585, 357, \dodoi{10.1038/s41586-020-2649-2}

\bibitem[{{Hunter}(2007)}]{matplotlib}
{Hunter}, J.~D. 2007, Computing in Science and Engineering, 9, 90, \dodoi{10.1109/MCSE.2007.55}

\bibitem[{{Ihara} {et~al.}(2007){Ihara}, {Ozaki}, {Doi}, {Shigeyama}, {Kashikawa}, {Komiyama}, \& {Hattori}}]{ihara2007}
{Ihara}, Y., {Ozaki}, J., {Doi}, M., {et~al.} 2007, \pasj, 59, 811, \dodoi{10.1093/pasj/59.4.811}

\bibitem[{{Justham}(2011)}]{justham2011}
{Justham}, S. 2011, \apjl, 730, L34, \dodoi{10.1088/2041-8205/730/2/L34}

\bibitem[{{Kerzendorf} {et~al.}(2018){Kerzendorf}, {Strampelli}, {Shen}, {Schwab}, {Pakmor}, {Do}, {Buchner}, \& {Rest}}]{kerzendorf2018}
{Kerzendorf}, W.~E., {Strampelli}, G., {Shen}, K.~J., {et~al.} 2018, \mnras, 479, 192, \dodoi{10.1093/mnras/sty1357}

\bibitem[{{Kerzendorf} {et~al.}(2013){Kerzendorf}, {Yong}, {Schmidt}, {Simon}, {Jeffery}, {Anderson}, {Podsiadlowski}, {Gal-Yam}, {Silverman}, {Filippenko}, {Nomoto}, {Murphy}, {Bessell}, {Venn}, \& {Foley}}]{kerzendorf2013}
{Kerzendorf}, W.~E., {Yong}, D., {Schmidt}, B.~P., {et~al.} 2013, \apj, 774, 99, \dodoi{10.1088/0004-637X/774/2/99}

\bibitem[{{Li} {et~al.}(2019){Li}, {Kerzendorf}, {Chu}, {Chen}, {Do}, {Gruendl}, {Holmes}, {Ishioka}, {Leibundgut}, {Pan}, {Ricker}, \& {Weisz}}]{li2019}
{Li}, C.-J., {Kerzendorf}, W.~E., {Chu}, Y.-H., {et~al.} 2019, \apj, 886, 99, \dodoi{10.3847/1538-4357/ab4a03}

\bibitem[{{Li} {et~al.}(2011){Li}, {Bloom}, {Podsiadlowski}, {Miller}, {Cenko}, {Jha}, {Sullivan}, {Howell}, {Nugent}, {Butler}, {Ofek}, {Kasliwal}, {Richards}, {Stockton}, {Shih}, {Bildsten}, {Shara}, {Bibby}, {Filippenko}, {Ganeshalingam}, {Silverman}, {Kulkarni}, {Law}, {Poznanski}, {Quimby}, {McCully}, {Patel}, {Maguire}, \& {Shen}}]{li2011}
{Li}, W., {Bloom}, J.~S., {Podsiadlowski}, P., {et~al.} 2011, \nat, 480, 348, \dodoi{10.1038/nature10646}

\bibitem[{{Liu} {et~al.}(2023){Liu}, {R{\"o}pke}, \& {Han}}]{liu2023}
{Liu}, Z.-W., {R{\"o}pke}, F.~K., \& {Han}, Z. 2023, Research in Astronomy and Astrophysics, 23, 082001, \dodoi{10.1088/1674-4527/acd89e}

\bibitem[{{Liu} {et~al.}(2022){Liu}, {R{\"o}pke}, \& {Zeng}}]{liu2022}
{Liu}, Z.-W., {R{\"o}pke}, F.~K., \& {Zeng}, Y. 2022, \apj, 928, 146, \dodoi{10.3847/1538-4357/ac5517}

\bibitem[{{Liu} {et~al.}(2021){Liu}, {R{\"o}pke}, {Zeng}, \& {Heger}}]{liu2021}
{Liu}, Z.-W., {R{\"o}pke}, F.~K., {Zeng}, Y., \& {Heger}, A. 2021, \aap, 654, A103, \dodoi{10.1051/0004-6361/202141518}

\bibitem[{{Liu} {et~al.}(2013){Liu}, {Pakmor}, {Seitenzahl}, {Hillebrandt}, {Kromer}, {R{\"o}pke}, {Edelmann}, {Taubenberger}, {Maeda}, {Wang}, \& {Han}}]{liu2013}
{Liu}, Z.-W., {Pakmor}, R., {Seitenzahl}, I.~R., {et~al.} 2013, \apj, 774, 37, \dodoi{10.1088/0004-637X/774/1/37}

\bibitem[{{Lundqvist} {et~al.}(2015){Lundqvist}, {Nyholm}, {Taddia}, {Sollerman}, {Johansson}, {Kozma}, {Lundqvist}, {Fransson}, {Garnavich}, {Kromer}, {Shappee}, \& {Goobar}}]{lundqvist2015}
{Lundqvist}, P., {Nyholm}, A., {Taddia}, F., {et~al.} 2015, \aap, 577, A39, \dodoi{10.1051/0004-6361/201525719}

\bibitem[{{Margutti} {et~al.}(2012){Margutti}, {Soderberg}, {Chomiuk}, {Chevalier}, {Hurley}, {Milisavljevic}, {Foley}, {Hughes}, {Slane}, {Fransson}, {Moe}, {Barthelmy}, {Boynton}, {Briggs}, {Connaughton}, {Costa}, {Cummings}, {Del Monte}, {Enos}, {Fellows}, {Feroci}, {Fukazawa}, {Gehrels}, {Goldsten}, {Golovin}, {Hanabata}, {Harshman}, {Krimm}, {Litvak}, {Makishima}, {Marisaldi}, {Mitrofanov}, {Murakami}, {Ohno}, {Palmer}, {Sanin}, {Starr}, {Svinkin}, {Takahashi}, {Tashiro}, {Terada}, \& {Yamaoka}}]{margutti2012}
{Margutti}, R., {Soderberg}, A.~M., {Chomiuk}, L., {et~al.} 2012, \apj, 751, 134, \dodoi{10.1088/0004-637X/751/2/134}

\bibitem[{{Marietta} {et~al.}(2000){Marietta}, {Burrows}, \& {Fryxell}}]{marietta2000}
{Marietta}, E., {Burrows}, A., \& {Fryxell}, B. 2000, \apjs, 128, 615, \dodoi{10.1086/313392}

\bibitem[{{Mattila} {et~al.}(2005){Mattila}, {Lundqvist}, {Sollerman}, {Kozma}, {Baron}, {Fransson}, {Leibundgut}, \& {Nomoto}}]{mattila2005}
{Mattila}, S., {Lundqvist}, P., {Sollerman}, J., {et~al.} 2005, \aap, 443, 649, \dodoi{10.1051/0004-6361:20052731}

\bibitem[{{Neunteufel} {et~al.}(2019){Neunteufel}, {Yoon}, \& {Langer}}]{neunteufel2019}
{Neunteufel}, P., {Yoon}, S.~C., \& {Langer}, N. 2019, \aap, 627, A14, \dodoi{10.1051/0004-6361/201935322}

\bibitem[{{Newville} {et~al.}(2014){Newville}, {Stensitzki}, {Allen}, \& {Ingargiola}}]{lmfit}
{Newville}, M., {Stensitzki}, T., {Allen}, D.~B., \& {Ingargiola}, A. 2014, {LMFIT: Non-Linear Least-Square Minimization and Curve-Fitting for Python}, 0.8.0, Zenodo,  Zenodo, \dodoi{10.5281/zenodo.11813}

\bibitem[{{Nomoto}(1982)}]{nomoto1982}
{Nomoto}, K. 1982, \apj, 253, 798, \dodoi{10.1086/159682}

\bibitem[{{Nugent} {et~al.}(2011){Nugent}, {Sullivan}, {Cenko}, {Thomas}, {Kasen}, {Howell}, {Bersier}, {Bloom}, {Kulkarni}, {Kandrashoff}, {Filippenko}, {Silverman}, {Marcy}, {Howard}, {Isaacson}, {Maguire}, {Suzuki}, {Tarlton}, {Pan}, {Bildsten}, {Fulton}, {Parrent}, {Sand}, {Podsiadlowski}, {Bianco}, {Dilday}, {Graham}, {Lyman}, {James}, {Kasliwal}, {Law}, {Quimby}, {Hook}, {Walker}, {Mazzali}, {Pian}, {Ofek}, {Gal-Yam}, \& {Poznanski}}]{nugent2011}
{Nugent}, P.~E., {Sullivan}, M., {Cenko}, S.~B., {et~al.} 2011, \nat, 480, 344, \dodoi{10.1038/nature10644}

\bibitem[{{Pagnotta} \& {Schaefer}(2015)}]{pagnotta2015}
{Pagnotta}, A., \& {Schaefer}, B.~E. 2015, \apj, 799, 101, \dodoi{10.1088/0004-637X/799/1/101}

\bibitem[{{Pan} {et~al.}(2010){Pan}, {Ricker}, \& {Taam}}]{pan2010}
{Pan}, K.-C., {Ricker}, P.~M., \& {Taam}, R.~E. 2010, \apj, 715, 78, \dodoi{10.1088/0004-637X/715/1/78}

\bibitem[{{Pan} {et~al.}(2012){Pan}, {Ricker}, \& {Taam}}]{pan2012b}
---. 2012, \apj, 760, 21, \dodoi{10.1088/0004-637X/760/1/21}

\bibitem[{{Pan} {et~al.}(2013){Pan}, {Ricker}, \& {Taam}}]{pan2013}
---. 2013, \apj, 773, 49, \dodoi{10.1088/0004-637X/773/1/49}

\bibitem[{{Patat} {et~al.}(2013){Patat}, {Cordiner}, {Cox}, {Anderson}, {Harutyunyan}, {Kotak}, {Palaversa}, {Stanishev}, {Tomasella}, {Benetti}, {Goobar}, {Pastorello}, \& {Sollerman}}]{patat2013}
{Patat}, F., {Cordiner}, M.~A., {Cox}, N.~L.~J., {et~al.} 2013, \aap, 549, A62, \dodoi{10.1051/0004-6361/201118556}

\bibitem[{{Pereira} {et~al.}(2013){Pereira}, {Thomas}, {Aldering}, {Antilogus}, {Baltay}, {Benitez-Herrera}, {Bongard}, {Buton}, {Canto}, {Cellier-Holzem}, {Chen}, {Childress}, {Chotard}, {Copin}, {Fakhouri}, {Fink}, {Fouchez}, {Gangler}, {Guy}, {Hillebrandt}, {Hsiao}, {Kerschhaggl}, {Kowalski}, {Kromer}, {Nordin}, {Nugent}, {Paech}, {Pain}, {P{\'e}contal}, {Perlmutter}, {Rabinowitz}, {Rigault}, {Runge}, {Saunders}, {Smadja}, {Tao}, {Taubenberger}, {Tilquin}, \& {Wu}}]{pereira2013}
{Pereira}, R., {Thomas}, R.~C., {Aldering}, G., {et~al.} 2013, \aap, 554, A27, \dodoi{10.1051/0004-6361/201221008}

\bibitem[{{Piersanti} {et~al.}(2014){Piersanti}, {Tornamb{\'e}}, \& {Yungelson}}]{piersanti2014}
{Piersanti}, L., {Tornamb{\'e}}, A., \& {Yungelson}, L.~R. 2014, \mnras, 445, 3239, \dodoi{10.1093/mnras/stu1885}

\bibitem[{{Podsiadlowski}(2003)}]{pod2003}
{Podsiadlowski}, P. 2003, arXiv e-prints, astro, \dodoi{10.48550/arXiv.astro-ph/0303660}

\bibitem[{{Polin} {et~al.}(2019){Polin}, {Nugent}, \& {Kasen}}]{polin2019}
{Polin}, A., {Nugent}, P., \& {Kasen}, D. 2019, \apj, 873, 84, \dodoi{10.3847/1538-4357/aafb6a}

\bibitem[{{Pradhan} {et~al.}(1996){Pradhan}, {Zhang}, {Nahar}, {Romano}, \& {Bautista}}]{iron1}
{Pradhan}, A.~K., {Zhang}, H.~L., {Nahar}, S.~N., {Romano}, P., \& {Bautista}, M.~A. 1996, in American Astronomical Society Meeting Abstracts, Vol. 189, American Astronomical Society Meeting Abstracts, 72.11

\bibitem[{{Rau} \& {Pan}(2022)}]{rau2022}
{Rau}, S.-J., \& {Pan}, K.-C. 2022, \apj, 933, 38, \dodoi{10.3847/1538-4357/ac7153}

\bibitem[{{Reback} {et~al.}(2022){Reback}, {jbrockmendel}, {McKinney}, {Van den Bossche}, {Augspurger}, {Roeschke}, {Hawkins}, {Cloud}, {gfyoung}, {Sinhrks}, {Hoefler}, {Klein}, {Petersen}, {Tratner}, {She}, {Ayd}, {Naveh}, {Darbyshire}, {Garcia}, {Shadrach}, {Schendel}, {Hayden}, {Saxton}, {Gorelli}, {Li}, {Zeitlin}, {Jancauskas}, {McMaster}, {W{\"o}rtwein}, \& {Battiston}}]{pandas}
{Reback}, J., {jbrockmendel}, {McKinney}, W., {et~al.} 2022, {pandas-dev/pandas: Pandas 1.4.2}, v1.4.2, Zenodo,  Zenodo, \dodoi{10.5281/zenodo.3509134}

\bibitem[{{Ruiz-Lapuente}(2019)}]{ruizlapuente2019}
{Ruiz-Lapuente}, P. 2019, \nar, 85, 101523, \dodoi{10.1016/j.newar.2019.101523}

\bibitem[{{Ruiz-Lapuente} {et~al.}(2004){Ruiz-Lapuente}, {Comeron}, {M{\'e}ndez}, {Canal}, {Smartt}, {Filippenko}, {Kurucz}, {Chornock}, {Foley}, {Stanishev}, \& {Ibata}}]{ruizlapuente2004}
{Ruiz-Lapuente}, P., {Comeron}, F., {M{\'e}ndez}, J., {et~al.} 2004, \nat, 431, 1069, \dodoi{10.1038/nature03006}

\bibitem[{{Schaefer} \& {Pagnotta}(2012)}]{schaefer2012}
{Schaefer}, B.~E., \& {Pagnotta}, A. 2012, \nat, 481, 164, \dodoi{10.1038/nature10692}

\bibitem[{{Schlafly} \& {Finkbeiner}(2011)}]{schlafly2011}
{Schlafly}, E.~F., \& {Finkbeiner}, D.~P. 2011, \apj, 737, 103, \dodoi{10.1088/0004-637X/737/2/103}

\bibitem[{{Shappee} {et~al.}(2013{\natexlab{a}}){Shappee}, {Kochanek}, \& {Stanek}}]{shappee2013}
{Shappee}, B.~J., {Kochanek}, C.~S., \& {Stanek}, K.~Z. 2013{\natexlab{a}}, \apj, 765, 150, \dodoi{10.1088/0004-637X/765/2/150}

\bibitem[{{Shappee} \& {Stanek}(2011)}]{shappee2011}
{Shappee}, B.~J., \& {Stanek}, K.~Z. 2011, \apj, 733, 124, \dodoi{10.1088/0004-637X/733/2/124}

\bibitem[{{Shappee} {et~al.}(2017){Shappee}, {Stanek}, {Kochanek}, \& {Garnavich}}]{shappee2017}
{Shappee}, B.~J., {Stanek}, K.~Z., {Kochanek}, C.~S., \& {Garnavich}, P.~M. 2017, \apj, 841, 48, \dodoi{10.3847/1538-4357/aa6eab}

\bibitem[{{Shappee} {et~al.}(2013{\natexlab{b}}){Shappee}, {Stanek}, {Pogge}, \& {Garnavich}}]{shappee2013a}
{Shappee}, B.~J., {Stanek}, K.~Z., {Pogge}, R.~W., \& {Garnavich}, P.~M. 2013{\natexlab{b}}, \apjl, 762, L5, \dodoi{10.1088/2041-8205/762/1/L5}

\bibitem[{{Shen} \& {Schwab}(2017)}]{shen2017}
{Shen}, K.~J., \& {Schwab}, J. 2017, \apj, 834, 180, \dodoi{10.3847/1538-4357/834/2/180}

\bibitem[{{Shen} {et~al.}(2018){Shen}, {Boubert}, {G{\"a}nsicke}, {Jha}, {Andrews}, {Chomiuk}, {Foley}, {Fraser}, {Gromadzki}, {Guillochon}, {Kotze}, {Maguire}, {Siebert}, {Smith}, {Strader}, {Badenes}, {Kerzendorf}, {Koester}, {Kromer}, {Miles}, {Pakmor}, {Schwab}, {Toloza}, {Toonen}, {Townsley}, \& {Williams}}]{shen2018}
{Shen}, K.~J., {Boubert}, D., {G{\"a}nsicke}, B.~T., {et~al.} 2018, \apj, 865, 15, \dodoi{10.3847/1538-4357/aad55b}

\bibitem[{{Shields} {et~al.}(2023){Shields}, {Arunachalam}, {Kerzendorf}, {Hughes}, {Biriouk}, {Monk}, \& {Buchner}}]{shields2023}
{Shields}, J.~V., {Arunachalam}, P., {Kerzendorf}, W., {et~al.} 2023, \apjl, 950, L10, \dodoi{10.3847/2041-8213/acd6a0}

\bibitem[{{STScI Development Team}(2018)}]{synphot}
{STScI Development Team}. 2018, {synphot: Synthetic photometry using Astropy}, Astrophysics Source Code Library, record ascl:1811.001.
\newblock \doeprint{1811.001}

\bibitem[{{STScI Development Team}(2020)}]{stsynphot}
---. 2020, {stsynphot: synphot for HST and JWST}, Astrophysics Source Code Library, record ascl:2010.003.
\newblock \doeprint{2010.003}

\bibitem[{{Taubenberger} {et~al.}(2015){Taubenberger}, {Elias-Rosa}, {Kerzendorf}, {Hachinger}, {Spyromilio}, {Fransson}, {Kromer}, {Ruiter}, {Seitenzahl}, {Benetti}, {Cappellaro}, {Pastorello}, {Turatto}, \& {Marchetti}}]{taubenberger2015}
{Taubenberger}, S., {Elias-Rosa}, N., {Kerzendorf}, W.~E., {et~al.} 2015, \mnras, 448, L48, \dodoi{10.1093/mnrasl/slu201}

\bibitem[{{Tucker} {et~al.}(2022{\natexlab{a}}){Tucker}, {Ashall}, {Shappee}, {Kochanek}, {Stanek}, \& {Garnavich}}]{tucker2022a}
{Tucker}, M.~A., {Ashall}, C., {Shappee}, B.~J., {et~al.} 2022{\natexlab{a}}, \apjl, 926, L25, \dodoi{10.3847/2041-8213/ac4fbd}

\bibitem[{{Tucker} {et~al.}(2022{\natexlab{b}}){Tucker}, {Shappee}, {Kochanek}, {Stanek}, {Ashall}, {Anand}, \& {Garnavich}}]{tucker2022b}
{Tucker}, M.~A., {Shappee}, B.~J., {Kochanek}, C.~S., {et~al.} 2022{\natexlab{b}}, \mnras, 517, 4119, \dodoi{10.1093/mnras/stac2873}

\bibitem[{{Virtanen} {et~al.}(2020){Virtanen}, {Gommers}, {Oliphant}, {Haberland}, {Reddy}, {Cournapeau}, {Burovski}, {Peterson}, {Weckesser}, {Bright}, {van der Walt}, {Brett}, {Wilson}, {Millman}, {Mayorov}, {Nelson}, {Jones}, {Kern}, {Larson}, {Carey}, {Polat}, {Feng}, {Moore}, {VanderPlas}, {Laxalde}, {Perktold}, {Cimrman}, {Henriksen}, {Quintero}, {Harris}, {Archibald}, {Ribeiro}, {Pedregosa}, {van Mulbregt}, \& {SciPy 1. 0 Contributors}}]{scipy}
{Virtanen}, P., {Gommers}, R., {Oliphant}, T.~E., {et~al.} 2020, Nature Methods, 17, 261, \dodoi{10.1038/s41592-019-0686-2}

\bibitem[{{Wang} {et~al.}(2009){Wang}, {Meng}, {Chen}, \& {Han}}]{wang2009}
{Wang}, B., {Meng}, X., {Chen}, X., \& {Han}, Z. 2009, \mnras, 395, 847, \dodoi{10.1111/j.1365-2966.2009.14545.x}

\bibitem[{{Wheeler} {et~al.}(1975){Wheeler}, {Lecar}, \& {McKee}}]{wheeler1975}
{Wheeler}, J.~C., {Lecar}, M., \& {McKee}, C.~F. 1975, \apj, 200, 145, \dodoi{10.1086/153771}

\bibitem[{{Whelan} \& {Iben}(1973)}]{whelan1973}
{Whelan}, J., \& {Iben}, Icko, J. 1973, \apj, 186, 1007, \dodoi{10.1086/152565}

\bibitem[{{Wu} {et~al.}(2017){Wu}, {Wang}, {Liu}, \& {Han}}]{wu2017}
{Wu}, C., {Wang}, B., {Liu}, D., \& {Han}, Z. 2017, \aap, 604, A31, \dodoi{10.1051/0004-6361/201630099}

\end{thebibliography}
\bibliographystyle{aasjournal}



\end{document}